# AI-Enhanced Deliberative Democracy and the Future of the Collective Will

## Abstract

This article unpacks the design choices behind longstanding and newly proposed computational frameworks aimed at finding common grounds across collective preferences and examines their potential future impacts, both technically and normatively. It begins by situating AI-assisted preference elicitation within the historical role of opinion polls, emphasizing that preferences are shaped by the decision-making context and are seldom objectively captured. With that caveat in mind, we explore AI-based democratic innovations as discovery tools for fostering reasonable representations of a collective will, sense-making, and agreement-seeking. At the same time, we caution against dangerously misguided uses, such as enabling binding decisions, fostering gradual disempowerment or post-rationalizing political outcomes.

## Authors:

**Manon Revel**[1], Meta, FAIR, New York (NYC), United States & Harvard University, Berkman Klein Center for Internet and Society, Cambridge (MA), United States. Orcid: 0000-0002-8335-946X. mrevel@mit.edu

**Théophile Pénigaud**, MacMillan Center, Yale University, Department of Political Science, New Haven (CT), United States. Orcid: 0000-0002-1760-8608. theophile.penigaud@yale.edu

[1] This project originated when Manon was at the Harvard Berkman Klein Center. Manon has since then moved to the Fundamental AI Research Lab at Meta.

## Acknowledgements:

We thank Shira Ahissar, Michiel Bakker, Léon Bottou, Paul Golz, Gillian Hadfield, Ariel Procaccia, Bernard Reber, Luke Thorburn, Pierre-Etienne Vandamme and Klaudia Wegschaider for feedback on this piece. We also thank the Oxford Institute for Ethics in AI, Caroline Green, Luise Muller and the participants of the 2025 Oxford Berlin Workshop on AI Ethics, as well as the members of the Normativity Lab for fruitful feedback on the ideas presented here.

# Content

> *And indeed, well if one day they really find the formula for all our wanting and caprices—that is, what they depend on, by precisely what law they occur, precisely how they spread, what they strive for in such-and-such case and so on and so forth; a real mathematical formula, that is—then perhaps man will immediately stop wanting [...]. Who wants to want according to a little table?*
>
> Fyodor Dostoevsky, *Notes from Underground* (1864)

# I - Introduction

The relationship between AI and democracy has primarily been discussed through the impact of algorithmically curated social media (e.g. Cohen & Fung 2021; Habermas 2023) and, more recently, the potential enhancements and threats generative AI poses to public discourse and trust (Kreps & Kriner 2023; Coeckelbergh 2023; Risse 2023; Summerfield et al. 2024). In this paper, we are interested in a different dimension of AI-driven democratic transformations, providing insight into the democratic potential and possible drawbacks of computational frameworks designed to assist citizens in coordination, deliberation, and consensus-building (Lazar, S., & Manuali, L. 2024).

Algorithms have long been employed to elicit, model, infer, and aggregate individual preferences across a wide range of applications. Over the past decade, heuristics have been further developed and deployed to algorithmically aggregate or synthesize these inferred preferences, purportedly producing a representation of the group's collective stances. Platforms such as Polis and Remesh have introduced frameworks for opinion mapping and consensus building, which have been applied in policy-making worldwide. More recently, language models (LMs) have been proposed as tools to generate collective stances on policy questions by predicting how individuals would evaluate various LM-generated collective statements, based on previously elicited information at the individual level (Fish et al. 2023; Tessler and Bakker et al. 2024). Related approaches have been put forward to simulate citizens' assemblies, that is, deliberation among a representative sample of citizens, and use their outcomes to align AI systems with human values (Mazeika, Mantas, et al 2025). Such AI-enabled representations of the collective preferences have no clear status, and their democratic function has yet to be clarified. These approaches certainly offer possibilities to enhance and scale representation and participation across different settings. They also raise novel questions related to their interface with existing democratic practices and norms.

While acknowledging the promise of these technologies, we emphasize the technical potentials and contingencies of design choices. The recognition that design choices are both necessary and consequential raises two imperatives. First, intelligibility: through a review of leading techniques that generate collective snapshots based on individual preferences, we make these choices explicit so their legitimacy can be assessed (and, if relevant, consented to), while also enabling the interrogation and revision of path-dependent design constraints. Second, conceptual clarity: drawing from democratic theory, we explore the extent to which AI-based democratic innovations are “deliberative” and politically “representative”. We view these methods as various ways to elicit and aggregate preferences and situate them within the historical role of opinion polls.

Human societies are poised to gain more information about themselves than ever before—a prospect that could prove either a blessing or a curse. This would be a curse if, following the logic of calculability—as evoked in our tribute to Dostoevsky in the epigraph—this increased knowledge of the social world served only to make political systems more reactive to immediate behaviors or preferences, ultimately eroding the ideal of collective self-government through public deliberation.

However, we argue that it is possible to harness this knowledge to deepen, rather than threaten, deliberative democracy. This possibility depends on taking seriously the design assumptions underlying current mathematical frameworks for capturing collective judgments (II), as well as their rightful status within the democratic system (III). We observe that there is an untapped potential to make these methods reflective and introduce the concept of AI Reflectors—defined as a model with reflective elicitation and argumentative syntheses (IV). We outline the technical and political uncertainties and challenges associated with the development of such tools (V) and highlight the narrow path toward this democratic future (VI).

# II - Statistical Methods and the Political Public Sphere

Democracy is the *power of the people, by the people, and for the people*.[2] However, determining what the people truly want is challenging. When can we confidently say that the people have spoken, that the state's actions reflect their will, or that political representatives are acting in their best interests? These questions are inherently elusive (Rosanvallon 1998), making democracy reliant on approximation tools, like polls and elections, to generate knowledge about the social world and the general will.

Hereafter, conceptualising the leading and emerging technologies’ design, we interrogate the kinds of information they provide about the social world and review the assumptions they are based on. We summarize in Figure 1 the core steps of each method and expand below on how they are operationalised in various systems.

[2] From U.S. President Abraham Lincoln’s famous Gettysburg Address (November 19, 1863).

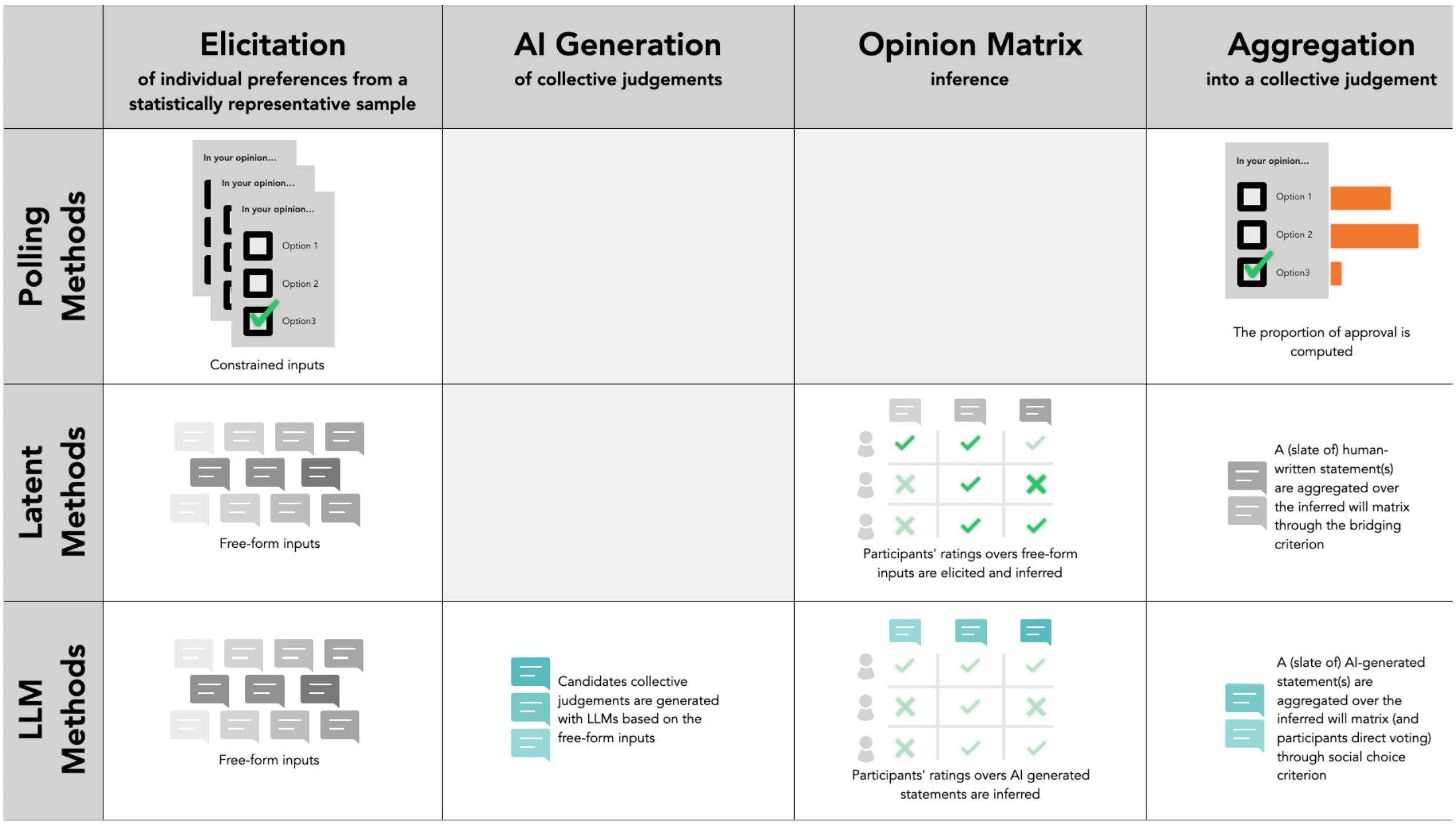

*Figure 1 – Overview of the key steps in the three generations of statistical methods for collective judgements (details are given below on how each method operationalizes these common steps)*

## *II.1 - Polling methods*

Polling has become integral to political landscapes, described by George Gallup as *The Pulse of Democracy* (Gallup and Rae 1940). Beyond predicting election outcomes, polls assess citizens' views on policy questions and inform campaign strategies (Hillygus 2011). They are widely influential to provide candidates with insights into their electoral chances and shifts in public opinion and are used to bring in citizens' insights into a wide range of policy decision-making (see, for instance, Colin Irwin's book [The People's Peace] for the use of polls in peace-making).

Scientific polling relies on the assumption that the opinion of a perfectly random sample can represent the entire population's public opinion, within a margin of error. In the absence of perfect sampling, pollsters use "quota-controlled surveys" (Hillygus, 2011) through stratified sampling (sampling independently or over-sampling subgroups of interest) or post-survey reweighting (weighing more individuals from groups predicted to have a low participation propensity).

Polling then juxtaposes readily-available opinions of a select few on a constrained set of prompts, providing a mathematically-mediated snapshot of unreflective, "raw" opinions that have not been mutually justified or critically examined. Such concerns led James S. Fishkin to introduce deliberative polling, a method designed to capture considered political opinions.

Classic polls capture preferences of individuals who may not be informed or have not engaged with others' opinions. Deliberative polls aim to replace this snapshot with a process where a cross-section of the population engages in genuine deliberations on a specific political question. This approach seeks to "get a picture of microcosmic deliberation: a representative mini-public of participants become informed as they weigh competing arguments on their merits" (Fishkin 2009, 54). Here again, the microcosm hypothesis is clearly stated "we can only know [what the people would think] if we start the deliberation with a good microcosm as representative as possible in both demographics and attitudes" (Fishkin 2018, 17).

The benefits are twofold: first, as in ordinary polling, the large number of participants can form a representative sample of the population, with quota corrections as needed.[3] Second, deliberation allows citizens to exchange views and modify their opinions based on relevant facts or arguments presented by others. Instead of the artificial aggregation of isolated opinions, deliberative polling provides a picture of the "considered judgement of the public"—or what the public *would think* if given a better opportunity to consider the issue (Fishkin 1995, 162). Deliberative polling involves polling participants before and after deliberation and learning phases, allowing citizens to form, inform, and update their opinions. This process highlights how opinions are constructed and change through a replicable exchange of reasons.

The argument for deliberative polls is fundamentally counterfactual: the "considered judgment of the public" reflects not the actual public will but a hypothetical one, assuming a more informed and deliberative public. Furthermore, deliberative polling still relies on a pre-defined restricted set of options that participants shall debate about and decide between. It involves aggregated reflective opinions after deliberation on a constrained set of options, with limitations regarding the agenda setting.

Overall, we may represent the polling methods in two simple steps:

1. Elicitation: Individuals (statistically representative of the whole) privately answer multiple-choice questions (before and after a deliberation in the case of deliberative polling).
2. Aggregation: The proportion of people that approved of each choice is computed.

The increasing use of polling in the 20th century and deliberative polling in the 21st century illustrates a method to capture social aspirations and its drawbacks. Most policy problems are open-ended, as are human thoughts, and the space of potential solutions is unbounded. We next explore technological approaches proposed to harness the open-endedness of human problems by aggregating open-ended statements into a collective outcome.

## *II.2 - Interlude: the opinion matrix*

While it is easy to aggregate approval votes on constrained options like in polls, the question of how to produce a collective outcome based on many free-form messages is much more tedious.[4] The solutions

[3] Note that while there are many people sampled, participants are broken down in small deliberation groups.

[4] We will also explore limitations of such tools and comment on the hardness to process open-endedness. "Polis has always faced a bottleneck in reporting, as conversations are labor-intensive to moderate effectively, and the results are often

to this problem typically rely on versions of the opinion matrix, also referred to as the will matrix by Konya et al. (2023).[5] The opinion matrix is composed of people's elicited (or, inferred) preferences over a list of free-form text (see Figure 1). These interactions between people and text are used to learn latent representation of both people and text that are used to select statements into an AI-mediated collective outcome.

Dating back to Charles E. Osgood (1957), social scientists have sought to capture human opinions indirectly by leveraging signals from a latent semantic space to model emotions and opinions. In the late 1970s, Jean-François Steiner developed a method called *semiometry*,[6] where participants rated their preference for 200-to-300 words on a 7-point scale. Each participant was represented by their preference profile over the words, while each word was characterized by its likability profile across participants. These person-word interactions were captured in a matrix format, enabling mathematical analysis.

A standard statistical method used in semiometry is Principal Component Analysis (PCA), which reduces the dimensionality of the interaction matrix. This means that if a voter is initially characterized by their 200-to-300 scores, the dimension reduction technique compresses these characteristics into a low-dimensional vector. PCA identifies a small number of latent dimensions—endogenously derived from the data—that explain most of the variance in word preferences. It then represents voters and words as functions of these few but meaningful latent dimensions. For example, the original semiometry analysis revealed a latent axis interpretable as duty versus pleasure. These latent dimensions provided insights into semantic relationships between words and social affinities among participants.

This approach of learning latent representations of users and items based on their interaction matrix (such as books on Amazon, movies on Netflix, or posts on Instagram) is now widely used to mediate interactions with information.

## *II.3 - Latent Methods*

Turning to latent methods: civic technologists have deployed techniques to understand citizens' perspectives on open-ended and emergent policy issues. A pioneer in this field is the platform Pol.is, launched by Colin Megill in 2014 and notably used in Taiwan to gather public input on regulations for Uber together with digital democracy leader Audrey Tang.[7]

information-dense." (quote from Colin Megil, CEO and founder of Pol.is sourced from Jigsaw's Medium post at https://medium.com/jigsaw/making-sense-of-large-scale-online-conversations-b153340bda55)

[5] Will matrix is defined there as "a matrix where every row corresponds to a human, and every column corresponds to an item containing information related to characteristics of potential futures" (Konya et al. 2023, 8).

[6] We thank Léon Bottou for pointing us to semiometry.

[7] See e.g., https://www.wired.com/story/taiwan-democracy-social-media/ and https://www.theguardian.com/world/2020/sep/27/taiwan-civic-hackers-polis-consensus-social-media-platform. Polis has been deployed around the world (e.g., https://www.rappler.com/philippines/pasig-city-uses-polis-consult-residents-open-streets-proposal/,

Pol.is allows users to submit short opinions on a given topic and encourages them to express preferences on others' opinions (stating whether they agree, disagree or pass). It creates a user-opinion matrix (similar to the collective will matrix, referred to as the polis opinion matrix), representing the vote of user *i* on opinion piece *j*. A common issue in such setups is that the collective will matrix is typically sparse—most people only vote on a small set of opinions, leaving their stance on most other statements unknown. In Pol.is, "missing values, corresponding to comments the participant in question did not see, are imputed by taking column-wise means of the non-missing values associated with the given comment" (Small et al. 2021). Next, the platform uses dimension reduction techniques, such as PCA (Hsiao et al. 2018), to learn a latent, low-dimensional representation of users' characteristics based on this inferred matrix of opinion.[8] It then clusters users based on these latent vectors, sorting them by similarity as implicitly defined by the clustering algorithm used.

Pol.is can directly identify consensual opinions from the opinion matrix—those with high approval across the board. It can also find opinions representative of the diverse comments. And, it derives metrics for opinion representativeness (where an opinion is considered representative if it is likely to be liked across distinct clusters).[9] The concept of selecting and ranking posts that receive high approval across diverse groups is often called bridging-based ranking (Ovadya and Thorburn 2022), and we refer to Pol.is' conceptualisation of representativeness as following the bridging criteria when aggregating statements into a winning outcome.

Another common solution to infer the unseen votes using statistical methods is to rely on the latent representation of users and comments derived from observed votes. An example of this approach is prominently used in Remesh, a collective response system used e.g., by the U.N. in peacebuilding contexts (Bilich et al. 2023) and to find common ground between Israeli and Palestinian peacebuilders (Konya et al 2025). In what follows, we outline the method used in this latter context.[10] Initially, Remesh has participants write opinions and vote on others' opinions. The resulting collective will matrix is sparse, as not all participants vote on all opinions. Remesh explicitly infers the unseen votes to complete the will matrix (based on their new paradigm, STUMP, that combines LMs and latent factors,

---

https://www.economist.com/open-future/2019/03/22/technology-and-political-will-can-create-better-governance) and inspired new governance approaches for AI as well (see https://time.com/6684266/openai-democracy-artificial-intelligence/).

[8] See https://compdemocracy.org/algorithms/.

[9] In more detail, Pol.is partition the users in the groups using clustering over the will matrix. It then defines *representativeness* of a comment c for a group g measures "how much more likely participants in group g are to place vote v on said comment than those outside group g" (https://compdemocracy.org/Representative-Comments/). The representativeness metric is defined as follows. Let g be a group of users and g' be the group of users including everyone but those in g. Let N(g) be the number of people in g who agree with comment c and T(g) the total number of people in g who have voted on comment c. Then, the probability that a person in g likes comment c is estimated as P(g) = (2+N(g))/(1+T(g)). Representativeness of c for g is then simply defined as the ratio P(g')/P(g). Next, *group-aware consensus* is defined as the product of P(g) across all groups g that partition the user-base. A high group-aware consensus score means that a comment was liked across groups and is designed to prevent the tyranny of the majority (https://compdemocracy.org/group-informed-consensus/).

[10] Remesh has used a variety of approaches and we report below the most recent one for its reliance on LMs described in November 2024 in https://www.youtube.com/watch?v=dTQl4midnvw.

(Konya et al. 2022), reportedly achieving an accuracy of 75 to 80%). A set of statements deemed representative is then selected, where representative posts are those that receive high, diverse approval across predefined demographic groups in a bridging fashion.[11] In Remesh's latest approach, these representative posts are parsed into an LM to generate distilled outcomes using few-shot learning.[12] Finally, experts review the LM output, and participants vote on each statement and rank them.

Another system of recent interest here is X's Community Notes algorithm. In this system, users can write fact-checking notes on any Twitter post and vote on others' notes (stating whether the note was helpful). The Community Notes algorithm then assumes a parametric form for the vote—that is, it assumes that each vote is the result of a simple mathematical operation—namely, the probability that person $u$ finds note $n$ helpful is assumed to be $v_{un} = m + i_u + i_n + f_u f_n$ where $m$ is a global constant, $i_u$ and $f_u$ (respectively, $i_n$ and $f_n$) are user-specific (respectively, note-specific) constants and $i_n$ is specifically the helpfulness score of note $n$. The algorithm uses the observed rating $v_{un}$ to run an optimisation framework and learn the parameters $m$, $i_u$, $i_n$, $f_u$, and $f_n$ that best approximate the observed votes.[13] These learned parameters are then used to select the notes with a high $i_n$ for they are "rated [as] helpful by raters with a diversity of viewpoint" where viewer diversity is estimated from the latent factors learned on past votes.[14]

Overall, we may represent the latent methods in three steps:

1. Elicitation: Individuals write free-form private messages.
2. Opinion Matrix Inference: Individuals vote on the free-form messages of others. Individuals are then represented and clustered based on these elicited interactions, and unobserved interactions are inferred.
3. Aggregation: The latent representations are aggregated into a collective outcome.

All these approaches rely on open-ended inputs, allowing participants to minimally engage with others' perspectives by voting on them and outputting a unique (or a set of) human-written opinions deemed most consensual, based on varied understandings of consensus. These approaches rest on open-ended statements and participants' votes which are combined into a matrix that serves as input for algorithms aimed at finding common ground. Winning statements are selected based on their ability to bridge across predefined (in Remesh) or endogenously learned (in Pol.is and Community Notes) groups. In its latest implementation, Remesh introduced the possibility to generate a synthesis of representative statements, leveraging LMs to generate open-ended outputs that may better represent a collective than a few selected participants' statements. We next explore technological approaches proposed to harness the open-endedness of LMs to generate coherent synthesis from open-ended collective judgments.

[11] Let statement A be approved by 80% of Democrats and 10% of Republicans and statement B be approved by 30% of Democrats and 35% of Republicans. Statement B is deemed more bridging because the smaller support across both political groups is greater. This approach has the potential to identify content cuts across e.g., political divide, but also bakes in important limitations as it may "inadvertently reduce the visibility of legitimate viewpoints" Revel et al (2025).

[12] Few-shot learning is a simple and common prompting technique with LM that shows an example of the task it asks to complete. For instance, a few-shot prompt would read: "1 + 1 = 2 and 3 + 4 = 7. What is 4 + 1 = ?"

[13] https://communitynotes.x.com/guide/en/under-the-hood/ranking-notes

[14] https://communitynotes.x.com/guide/en/under-the-hood/ranking-notes#determining-note-status-explanation-tags. See also https://jonathanwarden.com/understanding-community-notes/ for a deeper explanation of the algorithm.

## *II.4 - LM methods*

In two recent papers, Harvard's Generative Social Choice (Fish et al. 2023) and DeepMind's Habermas Machine (Tessler et al. 2024), scholars have developed generative machines that assume dynamic opinion formation in a boundless space of possibilities. In both cases, humans provide written inputs about a general topic during the elicitation stage. Specialized LMs are used to generate statements representative of the group. The liking of each individual for each representative statement is then computed using (potentially other) LMs. Even if the heuristics do not directly elicit an opinion matrix, they infer one based on free-form opinions and demographics. Finally, the predicted preferences are aggregated to maximize some notion of social welfare, selecting the winning representative statement.

The Habermas Machine introduced an LM-based approach to accelerate consensus finding across a plurality of political opinions by modeling humans' liking of a variety of algorithmically generated consensus statements. Harnessing the power of LMs to handle and produce free-form statements, the Habermas Machine takes as input a sample of free-form reflective opinions written by participants on stated issues and outputs an outcome deemed consensual.

The Habermas Machine synthesizes open-ended inputs from humans through the following steps:

1. Elicitation: Individuals privately write open-ended answers to open-ended questions.
2. Collective judgements generation: A set of candidate consensus statements is generated by an LM.[15]
3. Opinion matrix inference: The preferences of each candidate for each statement from the candidate set in 2 are inferred as follows. First, in a preliminary stage, each participant provides a ranking over a set of statements (on an issue unrelated with that of the main task). For each participant, a personalized reward model (PRM)[16] is trained based on these rankings. Specifically, the PRM learns to predict a number corresponding to how much a person with a given preference statement will like each consensus statement from the candidate set. That is, an opinion matrix, that assigns a score for each person-consensus statement pair, is inferred by the PRM. For each participant, the PRM scores are turned into a ranking over the candidate set.

[15] Note that the Habermas Machine could be prompted to generate various kinds of outputs—while it is asked to generate a consensus statement in the current approach, the prompt could be tuned to highlight, e.g., areas of dissensus.

[16] A reward model is an LM that is adapted to take as an input a pair of text and to produce a scalar score. In the case of the Habermas Machine, the personalized reward models are trained to learn the ranked preferences of humans collected during training phases.

4. Aggregation: The ranking profiles over the consensus statements are then aggregated into a single winner using the Schulze method.[17] This procedure is repeated 4 times so that 4 winning statements are selected and sent back to the participants. Participants rank the 4 top consensus statements selected in Step 4. The Schulze method is used again to select a winning statement, which is privately sent back to all participants.
5. Elicitation with reflection: Individuals write revisions of the winning statements. Steps 2-4 are repeated to select the final collective judgment.

Note that the Habermas Machine infers an opinion matrix instead of eliciting it directly. It uses a social choice method to aggregate inferred preferences—namely the Schulze method. The authors explain that the Schulze method is used (Tessler et al. 2024, SM 38) for its independence of clone property (that ensures that statements with similar preference ratings are not jointly selected) enforcing some notions of balanced representation.

Generative Social Choice (Fish et al. 2023) is an approach designed to generate a slate of *k* statements (as opposed to a consensus statement) that represent the diverse perspectives of a group. Here's a high-level overview of how it works:

1. Elicitation: Individuals privately write open-ended answers to open-ended questions.
2. Pre-processing: Individual statements are summarized in fifty items and an LM infers whether each participant would agree with each of the fifty items. Participants are then clustered based on their inferred approval of the discretized items.
3. Collective judgements generation: An LM generates a statement based on these individual inputs. The LM is prompted to generate a query that maximizes the level of satisfaction of the *n/k*-th person (where individuals are ordered by their satisfaction level for the generated statement).[18]
4. Opinion matrix inference: The LM then simultaneously generates statements and predicts the opinion matrix (which represents the support each participant would give to the generated statement). More precisely, an LM is given as an input survey answer from 1. The LM then outputs the rating that individual would give to the generated statement.[19] The group formed by the *n/k* persons that most supports the generated statement are labeled as satisfied.
5. Aggregation: The generated statement is added to the set of output statements.
6. The opinions of the individuals not yet satisfied in 4. are sent back into an LM and steps 3-5 are repeated until there are *k* statements in the slate.

[17] https://en.wikipedia.org/wiki/Schulze_method
[18] Because the LM did not appear as sensible as needed to the n/k value, the authors developed an ensemble method through which they generated queries based on clusters of opinions or random draws of opinions.
[19] More precisely, the authors use the token probabilities to construct a probability distribution over the inferred rating.

This method aims to capture a range of perspectives by generating then selecting statements that reflect various groups deserving representation, ensuring that the final set of statements represents plural viewpoints. Specifically, the authors prove that their method satisfies a social theoretic axiom called *justified representation*. In a context where *k* items are to be selected, justified representation ensures that any group of size at least *n/k* with cohesive preferences is minimally represented.[20] Crucially, note that Generative Social Choice is the only method thus far providing verifiable theoretical guarantees in terms of representation: the slates generated by this method are guaranteed to provide *justified representation* (assuming that the LMs perform its task of predicting utilities correctly).

Both latent and LM methods involve inputting open-ended thoughts that are aggregated through the uninterpretable inference of a will matrix.[21] The Habermas Machine and Generative Social Choice are similar to the methods used in collective response systems in that they take open-ended thoughts as input and rely on an (inferred) will matrix. However, they differ in that both heuristics explicitly harness the unbounded structure of potential outcomes by leveraging the generative and summarisation power of LMs. Unlike the collective response systems, these methods do not rely on bridging methods (Ovadya and Thorburn 2022) for aggregation. Instead, they use social choice theoretic tools, such as the Schulze method and the justified representation axiom, to guide the model generation towards representative statements. In the case of the Habermas Machine, participants also have the opportunity to reflect and update their thoughts.

Thus far, we explored old and new instruments designed to capture the "collective will", among which opinion matrix and LM methods. In what follows, we will review how these methods provide additional information about the collective will and question their place within the democratic process. From there, two questions arise: What is the status of this new information about the collective will? (III) And what are its distinctive features? (IV)

---

[20] Technically, justified representation is defined for approval preference profile (where users either approve or disapprove a statement). The axiom reads: a committee of k items satisfies justified representation if for every group of size at least n/k in which the members approve of at least one item in common, there is at least one group member that approves of an item in the committee. In Generative Social Choice, the authors generalize the setting to numeric ratings and strengthen this axiom. The natural generalization of justified representation requires that there shall not exist any group of size n/k in which every member rates an item not in the committee higher than all the items in the committee. They further strengthen the axiom by matching statements with equal-sized blocks of participants and guaranteeing that no *n/k* agents can find a statement that they all prefer to their respective matched statement. They call this new axiom *balanced justified representation*. It implies justified representation and conceptually ensures that a group much larger than *n/k* cannot be represented by a single statement.

[21] Note that considerable implementation details are not discussed here: the reader may refer to the Habermas Machine and the Generative Social Choice supplemental material for more information.

# III - Democratic Theory and AI-Enhanced Deliberative Innovations

Our central claim in this section is that AI-facilitated collective judgments may, at best, foster collective reflectivity within the political public sphere—much like polls, though possibly with far greater scope and depth. At worst, they risk being used to bypass public deliberation by allegedly simulating it. Either way, they are likely to be used in support of political decisions, which calls for caution (V).

## *III. 1 - The indeterminacy of the will*

Let us begin by addressing a widespread, in our eyes misguided view about democracy, that we wish to make clear we do not endorse. According to this view, democracy is essentially defined by congruence—that is, alignment with the population's current preferences. In contrast, following a well-established line of democratic theory, we take the "collective will" to be essentially indeterminate (Rosanvallon 1998): there is no way that preference elicitation can result in the unambiguous expression of the collective will.

*Empirical indeterminacy:* The very notion of "preference" is ambiguous, as it may denote (*inter alia*) what an individual prefers absolutely; what they prefer against specific alternatives (or every relevant alternatives); what they prefer consistently over time; what they prefer after careful reflection; or what is conducive to their welfare, regardless of avowed preferences.

Moreover, expressed preferences frequently diverge from latent ones, and the process of eliciting them can itself reshape them. Expressed preferences are often mediated by the perceived structure of available options (Elster 1986). Riker (1988) and Sunstein (1991) take this argument further, suggesting that preferences are never merely revealed but are always partially constructed within specific decision contexts.

*Aggregative indeterminacy:* Even if you were certain about what is to be captured exactly, social choice theory has long shown that collective outcomes—assuming well-defined individual preferences—often lack consistency across different aggregation rules. While social choice theory has provided crucial insights in the guarantees and challenges of various aggregation rules, the latter may not, alone, suffice to justify the validity of a collective outcome.

*Normative indeterminacy:* Most importantly, the collective will is indeterminate *de jure* before it proves to be so *de facto*. As per Claude Lefort's famous definition:

> *[D]emocracy inaugurates the experience of an ungraspable, uncontrollable society in which the people will be said to be sovereign, of course, but whose identity will constantly be open to question, whose identity will remain latent.* (Lefort 1986, 303—304).

If the collective will were readily determined, no space would remain for collective self-determination. Personal autonomy is not merely a matter of acting on accurate knowledge of one's putative interests; rather, it consists in deciding what one wants for oneself, free from coercion and deception. To be sure, extreme ignorance about oneself—or about the means to achieve one's ends—undermines autonomy. A minimal degree of self-knowledge and awareness of the context of choice is required for an act to count as self-directed. But autonomy also includes the freedom to choose *against* one's best interests. Hence, the Dostoevskian outcry that serves as our epigraph: "who wants to want according to a little table?"

By the same token, democracy cannot be reduced to *congruence*—the perfect mapping of policies onto pre-existing preferences. Democracy is equally, and perhaps more deeply, about sustaining the conditions under which the people can constantly revisit who they are and what they value. It has to do with the conditions under which opinions are formed, expressed, and challenged in light of new information and competing claims. The task is not to interpret citizens' preferences, but to transform them—in this instance, to allow them to evolve within procedurally fair settings, under the best possible conditions, attention being given to power relations.

Some social engineers and political theorists tend to conflate democracy with congruence by promoting democratic innovations aimed at tracking the collective preferences in order to circumvent the slow, participatory process of collective self-determination. Let's call this view the calculability hypothesis.

*Calculability hypothesis*: preferences exist independently, as standalone realities, or latent variables. The only question, therefore, is how best to compute those preferences to obtain a representative outcome (variably understood).

Recent proposals in the field of AI "alignment" illustrate this risk. The issue of "deliberative" or "democratic" alignment is typically approached through the lens of aligning AI agents with responses elicited from an inclusive—ideally global—sample of the population (Edelman & Klingefjord 2023 ; Konya, Turan, Ovadya et al. 2023 ; Mazeika, Mantas, et al 2025). We are concerned that this may introduce a technological shortcut, whereby the democratic process is bypassed in favor of a procedure that purports to simulate or substitute for it. Such approaches, we contend, tend to gloss over the fundamental requirement of collective agency. There is a crucial difference between, on the one hand, involving the public by eliciting its current preferences to inform the design of agentic AI, and, on the other hand, subjecting that design—together with its purposes and its broader ethical, political, and economic implications—to open democratic contestation. Only the latter preserves democracy's core commitment to collective self-determination.

## *III. 2 - The Two Meanings of Representation : the Portrait and the Mirror*

How can one assert that democratic institutions *must track* the people's will—a requirement essential for any political system to be deemed a democracy—while also claiming that this collective will can never be definitively located[22]? This seeming paradox dissolves when the democratic process is conceived of as forward-looking rather than as processing a fixed sum of information. In the latter case, the best "representation" of the people's preferences equate with the best aggregation rule and elicitation methods. The most accurate the elicitation, the fairer and most sensitive to individual preferences change, the best the representation. Representation, in this view, is pictorial in nature, and as such, static. Just as a portrait fixes what a person once was, retrospective representation fixes what a collective—here, the people—has been, with the style of the portrait being analogous to the aggregation rule (for example, justified representation).

*Political* representation, by contrast, is *dynamic* and future-oriented. It functions as a mirror. Elected representatives serve as the medium through which the people continually reflect on and reconsider their identity, expectations, and conceptions of the common good (Cohen 1989; Ubinati 2006; Mansbridge 2019). As highlighted by Manin:

> *Both popular self-government and absolute representation result in the abolition of the gap between those who govern and those who are governed, the former because it turns the governed into the governors, the latter because it substitutes representatives for those who are represented. A representative government, on the other hand, preserves that gap.* (Manin 2010, 174-5)

Electoral representation *presupposes* that the people exist substantively beyond the actions of their representative[23]. No elected representative individually, nor Parliament collectively, nor the democratic system in its entirety can ever claim to embody the will of the people. Rather, essential to the democratic process is what Urbinati (2014) describes as the *indirectness* of politics—the reciprocal representation of citizens' judgements by their representatives and vice versa—enabling the demos to engage in self-reflection and evaluate its laws, institutions, and leaders.

The dream of reaching the heart of the general will has haunted all technical efforts at preference elicitation since the early days of polling: "owing to polls, it was said, it would at last be possible to find out what people truly and spontaneously believe or want, without any adulterating mediation" (Manin, 2010, 172). In fact, "opinion polls, like demonstrations and petitions, do not deliver the pure, undistorted opinion of the public" (173). These are all "partial" manifestations of it. These "representations" are not absolute, but they all constrain to some degree what can be said about the people's will. They prevent a monopoly of political representatives over interpreting the general will by allowing the people to "reveal themselves as a political entity capable of speaking apart from those who govern" (174). Polls have "predictive value and bring to light significant cleavage" (230). Furthermore, they “lower the cost of individual political expression” particularly “because they are peaceful, whereas demonstrations often carry the risk of violence” (231).

[22] The "proper relationship between citizen preferences and the laws that govern them" is the "central normative problem" of democracy" (Rehfeld, 214).

[23] This opposition has its intellectual roots in Hannah Pitkin's distinction between "acting for" and "standing for" (Pitkin 1967). The former implies acting on behalf of the constituents while remaining *responsive* to them whereas the latter involves substituting for the represented, with the risk of shortcuts (Lafont 2019).

It thus appears that representations of the collective will, as forms of collective *self-probing*, are not to be dismissed, provided they remain reasonable, reflective, and prospective[24] . For both empirical and normative reasons studied above, reasonable representations remain necessarily *partial*, while potentially deepening deliberative democracy by fostering ongoing debates on the people’s identity and future[25].

## *III. 3 - The Two-track Model of Deliberative Democracy*

It follows from this clarificatory section that improving democracy essentially means equipping the people with the means to better understand the system and to act more autonomously within it, in order to hold it accountable to their interests or enlightened preferences. Just like polls, AI-based deliberative innovations may allow both leaders and citizens to articulate their views and influence the political process, making it *more self-conscious and enhancing collective reflexivity and agency*—they are self-governance compasses.

Just because the general will is indeterminate does not mean that no improvement is possible in terms of citizens’ autonomy, equality, or agency, or in their capacity to hold their representatives meaningfully accountable to their (ever-evolving and self-critical) collective will. Having a plurality of tools to tap into institutionalized and grassroot claims—that are nonetheless never absolutely representative—has been recognized central to modern politics (Ganz 2001).

Habermas famously theorized deliberative politics as proceeding "along two tracks that are at different levels of opinion- and will- formation, the one constitutional, the other informal" (Jurgen Habermas 1998, 314). The unregulated public sphere forms a "context of discovery," fostering debate and opinion formation, while formal institutions serve as a "context of justification" (307) converting public opinion into binding decisions and accounting for them to the citizenry. As long as public discourse informs institutions and institutional decisions remain transparent and accountable to citizens regarded as political equals, political decisions can be deemed legitimate, albeit fallible. This implies, in turn, that citizens be given multiple opportunities—enshrined in the rule of law—to contest or revisit earlier decisions in light of new information and claims made by fellow citizens.

[24] We introduce the notion of "reasonable representations" to emphasize three key features: 1. **Non-arbitrariness** These representations are not arbitrary, as they respond to independent signals and remain sensitive to real changes in individuals and collective preferences. 2. **Interpretative Function** They provide interpretations of the collective will, enabling the citizenry to reflect on itself—much as electoral representation does. 3. **Inherent Fallibility** They are fallible: although they attempt to capture something true about the collective will, they should not be mistaken for an unadulterated expression of it.

[25] If the general will is intractable, one might then ask why voting should authorize certain individuals to make binding decisions. Certainly, voting is a “blunt instrument” (Chapman 2022, 218). However, voting makes citizens’ equality concrete and transparent. Elections ensure that those who are delegated the power to decide on behalf of the community depend on a simple, accessible mechanism, the stakes of which can be clearly identified (30-41). We fail to see how even the highest levels in AI literacy could meet the same standard of aggregative equality.

Some scholars have argued that the discovery function of the public sphere has been jeopardized in recent years (Bednar 2021), with rising affective polarization (Boxell 2024), "islands of communication" (Habermas 2023, 42, 44) and what Henry Farrell calls "publics with malformed collective understandings."[26] We suggest turning the problem on its head and explore how AI-enhanced tools for self-discovery may empower the citizenry to be more agentic on the Habermasian second track.

In what follows, we consider AI-powered deliberative innovations as additional discovery tools "to mobilize and pool relevant issues and required information" (Habermas 2006, 416).

*Agenda-setting:* One potential function of AI-facilitated collective judgements could be to bring forward aspirations that have not received, or have not yet received, sufficient resonance within the formal political sphere. Because of their open-ended and lesser-framed nature, AI facilitated collective judgements may surface hitherto unarticulated grievances and uncrystallized issues. Interestingly, some studies started leveraging the summarization capabilities of LMs to assess the extent to which citizens' freely expressed aspirations—those proposals they care about most—are actually reflected in congressional policymaking.[27]

*Synergies with other democratic innovations:* Another potential function of AI-facilitated collective judgements could be to help bridge the gap between institutions such as Parliament or Citizens' Assemblies, on the one hand, and the broader public on the other. Petitioning has long been recognized as a fundamental democratic right, as evidenced by its enshrinement in the First Amendment—a testament to its foundational role in political modernity. While not binding, AI facilitated collective judgements could nonetheless serve as innovative, bottom-up mechanisms for extending qualitative deliberation to wider publics. It also may be implemented *within* citizens' assemblies to help participants cluster arguments and find common grounds (Landemore 2023).

In addition to these functions, in the next section, we explore a more original possibility opened by digital democracy through the application of artificial intelligence in forming collective judgments. "AI Reflectors," as we term them, may reduce the costs associated with democratic deliberation itself, without diluting its substance or representing a shortcut. Indeed, deliberation demands considerable investment in time, cognitive effort, and emotional resources. This requirement partly explains why citizens delegate the handling of their conflicting interests to representatives (Mansbridge 1980; Mutz 2006). By generating a dynamic yet accessible mapping of opinions and reasoning, and by highlighting areas of consensus and dissensus that necessitate either deeper exploration or a shift toward compromise, AI reflectors could grant citizens accelerated access to their differences and disagreements.

[26] https://www.programmablemutter.com/p/were-getting-the-social-media-crisis. For a balanced review of the possible impact of social media on deliberative democracy, see (Cohen and Fung 2021).

[27] See e.g. Velez, Yamil and Ewig, Alec, "Is Congress Listening? An Empirical Test of the Second Face of Power Using Tailored Surveys", draft paper.

# IV- The case for AI Reflectors

## *IV.1 - Blueprint*

Most of the techniques listed in section II prove heavily aggregative: they infer preferences, they are not intended to transform them. Such calculability-driven approaches tend to infer or elicit instinctive preferences, unreflected behaviors or static demographics. The Habermas Machine represents a first step toward a reflectivity-focused process.

Inspired by the Habermas Machine and the Generative Social Choice framework and by the earlier methods such as that of Pol.is and Remesh, we introduce AI Reflectors as a *technique* designed to (i) process open-endedness, (ii) elicit reflective preferences, and (iii) synthesize collective judgements (e.g., areas of consensus and irreconcilable pluralism) and as a *process* serving as a self-discovery tool for the citizenry to reflect on itself.

*Open-endedness*: Most socio-political problems are inherently ill-defined and open-ended; humans rarely consider them in the format of multiple-choice questions. To minimize the impact of question framing on elicited preferences, AI Reflectors take open-ended perspectives as input.

*Reflective Preferences*: Humans often lack readily available informed opinions (Lichtenstein and Slovic 2006). Preferences revealed through polls or interactions with online systems can conflate many desiderata (Anderson 2001; Thorburn 2022), and factual beliefs have been found to increasingly diverge along partisan lines (Strömbäck et al. 2024; Lichtenstein and Slovic 2006) which highlight the need for "preference construction." Deliberative democrats have proposed the *exchange of reasons* through deliberation as a means to "counteract the pernicious fragmentation of the public sphere [...] and facilitate the comprehension of choices" (Manin 2017) reach mutual understanding (Gutman and Thompson 1998) or rationally-justified outcome (Habermas 1998).

*Generate Consensus and Constructive Dissensus*: Historically, making sense of open-ended text has been a challenging task. Until the early 2020s, leading methods in Natural Language Processing (NLP) for summarizing text, known as extractive methods, focused on concatenating statistically important sentences (El-Kassas et al., 2021). While these methods helped reduce large corpora of text, they were "far away from the human-generated summaries" (El-Kassas et al., 2021). Identifying nuanced trends in open-ended corpora—such as areas of consensus and reasons for dissensus—seemed impossible until the introduction of LMs (Zhang et al. 2024). With LMs, there may no longer be a need to infer the will matrix or to arbitrate between different bridging hypotheses. While these heuristics make the system more interpretable, we explore the potential of LMs to outperform such ad hoc procedures in specifically summarizing consensus and pluralism in a collection of reflective opinions.[28]

In turn, an AI Reflector would rely at minimum on two steps:

1. Reflective Elicitation: Humans privately write open-ended answers to open-ended questions. The LM prompts the human to reflect on their thoughts and may provide experiences, testimonies, arguments, and contradictory facts as needed or if requested by the user.
2. Synthesis Generation: An LM generates a collective judgment based on all the reflective open-ended thoughts of the humans (intermediary steps may include inferring will matrices, or providing discursive analyses of the reasons and emotions stated with a given opinion by individuals as is further discussed in the last section).

AI reflectors are meant to provide a reasonable representation of itself to the citizenry while letting individuals reflect on their wants in the context of that of others. They first elicit reflective preferences and second synthesize those into a representative statement. They mediate, in turn, information both at the individual and collective level and shall deserve intense scrutiny—normatively and empirically.

## *IV.2 - Augmenting Deliberation?*

Introducing AI-facilitated techniques (e.g., the Habermas Machine) as a new deliberative arena may understandably prompt pushback. Some might argue that a collective judgment cannot be considered deliberative without binding outcomes, as deliberation involves weighing the pros and cons of a course of public *action* (Manin 1987; Mansbridge 2015; Chambers and Warren 2023). Others may express concerns about a process where participants never engage face-to-face: how can participants be made aware of and accountable to others' views and situations (Fishkin 2025)? Moreover, democratic deliberation is generally conceived of as persuasion-oriented; it involves providing justifications for our views that we believe others might endorse (Rawls 1993), remaining open to objections, and seeking rationally-motivated collective agreement (Habermas 1998). At first glance, AI reflectors enable reflexivity, or internal deliberation, not mutual persuasion (but this theoretical feature requires much empirical testing).

Nevertheless, AI Reflectors may represent a deliberative update in two senses. First, we preserve from deliberative democracy the minimal requirement of taking *uncoerced*, *informed*, and *other-regarding* contributions as *inputs*. Second, the output is not treated as the definitive expression of a collective stance, but rather as something to be recursively reclaimed, revisited, and discussed—both by participants, stakeholders or observers. Depending on certain conceptual choices, AI reflectors may generate a signal characterized by several unique features, at the input, throughput, and output levels[29].

[28] While explicit aggregation tools (e.g., bridging, social welfare functions, or justified representation) enforce an interpretable account of representation and provide some level of interpretability, zero-shot prompting can also technically produce group statements. Note, in fact, that Fish et al. (2023) also ask participants to rate output statements generated directly by an LM. Participants express a similar level of liking for these zero-shot outputs as for those produced by the more sophisticated method.

*Input – self-initiated, reflective contributions:* A first feature of the input is that AI Reflectors enable unconstrained, open-ended judgements. Agonistic democrats (Laclau and Mouffe 2014) and political constructivists (Disch 2021) have emphasized that no cleavage is neutral. Likewise, critics from sociology have noted that framing gives pollsters a questionable role in shaping opinion by deciding which issues are worth surveying and creating a false impression of knowledge about the social world, where responses are in fact heavily solicited and often lack consistency (Converse 1974; Bourdieu 1984).

In contrast, empowering people arguably involves granting them more autonomy both in question-framing and problem-solving. In this respect, the unconstrained, AI-facilitated approach to public issues holds potential. Given the unconstrained format of the input, the synthesis is more likely to reveal whether the dominant framing of political problems is legitimate or problematically distorted. As long highlighted by Habermas, there is a risk that "as parties become arms of the state, political will-formation shifts into a political system that is largely self-programming" (Jürgen Habermas 1997, 52). Two different things are "mass loyalty extraction" on the one hand and the public taking "responsibility for the pool of reasons that the administrative power can handle instrumentally, but cannot ignore" on the other hand (59).

*Throughput – equalizing information and expression:* AI Reflectors aim to foster and elicit informed, other-regarding, well-considered judgments. LMs might serve as inner assistants, incentivizing what Robert Goodin termed "deliberation within" (Goodin 2000). Some studies suggest that LMs can be used to enhance reflectivity (Duelen, Jennes, and Van den Broeck 2024; Hung et al. 2024). Ideally, LMs could assist users by providing high-quality information, asking for clarifications or justifications for their judgments, and presenting contradictory facts, arguments, stories, and testimonies.

Another interesting feature of AI Reflectors is their ability to meet the demand for input inclusivity, transcending language barriers and mitigating distortions caused by power relations. In contrast to face-to-face deliberation, there is no risk of capture by over-involved participants or undue influence from social authority or rhetorical skills. From this perspective, AI Reflectors may appear to fulfill Rousseau's dream of harmonious collective deliberations—freed from rhetorical manipulation and factional influence—where "every citizen states only his own opinion" (Rousseau 1997, 60).

*Output – collective self-knowledge and the rebalancing of power relations:* While we can envision letting a large number of individuals input a statement into an AI Reflectors, much technical research is needed to guarantee that each statement is equally attended to. Regardless, just as AI-generated content derives its value from the judgments *we* form about it—since only humans, as beings for whom outcomes matter, create and assign values—AI-generated collective stances derive their democratic authority from the sense *we* (critically) make of them. Now, how is this collective stance to be unpacked? And what special information can it be expected to convey?

[29] The input/throughput/output distinction—commonly used in legitimacy theory (Bekkers & Edwards, 2007; Schmitt 2013) and later applied to the evaluation of democratic innovations (e.g. Caluwaerts and Reuchamps, 2015; Reuchamps and Suiter, 2016)—refers respectively to the sources of participation and demands (input), the quality of internal processes and decision-making (throughput), and the resulting outcomes or policies (output).

First, AI reflectors may help people identify common ground from which to articulate and strengthen legitimate grievances and concerns. Second, they could help make agenda-setting more bottom-up, highlighting appropriate options between which to organize debate or decision-making[30].

Finally, AI reflectors may help the citizenry overcome collective action problems and deliberative bottlenecks. As democratic skeptics have noted, deliberative processes are often structurally undermined by disparities in information access, persuasive capacity, and cognitive resources. deliberation can end up reinforcing outcomes favored by those already in power, since belief equilibria are endogenous to the political system (Przeworski 1998). By enabling citizens to test the convergence of their core preferences, AI-facilitated judgments could foster a deeper collective self-awareness, empowering self-confident, enduring collective actions.

Traditional approaches to democratic affordances have primarily focused on bridging the gap between representatives and those they represent. Through emerging participatory platforms, digital democracy would both allow citizens to communicate their interests to their representatives, enable those representatives to monitor the opinions and preferences of their constituents, and facilitate citizen involvement in the co-construction of public policies (Deseriis, 2021; Small et al. 2021). In contrast, AI Reflectors are not (primarily) aimed to enable citizens to self-govern, as in models of disruptive direct democracy, nor to empower political elites to more effectively capture and navigate public expectations. Rather, they are intended to feed the conversation *among* citizens—to help individuals better understand what they want and where they stand in relation to others. By doing so, AI Reflectors could support the emergence of a renewed political agenda: one still enacted by traditional institutions, but subject to oversight by a citizenry rendered more vigilant and resistant to the dominant plebiscitary and adversarial rhetoric of electoral politics.

# V - Building AI Reflectors

Building on the last sections, we are interested in the design of an AI tool that can elicit reflective thoughts at scale and generate meaningful collective judgments. The advent of LMs, capable of processing and generating open-ended text, represents an epistemological rupture in the field of modeling the social world. Nonetheless, the viability of AI Reflectors depends on the capacity of an LM to produce valuable summaries not spoiled by biases, hallucinations or coercive persuasion. While one takes a leap-of-faith when using an LM, we review essential challenges and limitations below. We shall care about such a system's scalability as much as its meaningfulness (is the outcome sensible or what are its side effects) and simplicity (perhaps the most overlooked democratic desiderata in the technology world).

[30] We thank Philip Pettit for suggesting this function.

## *V.1 - Designing for scale*

First of all, we observe that current LM-based solutions for collective sense-making are difficult to scale—while scale being a prominent reason to use them in a deliberative context (Landemore 2023). How can we achieve scalability without sacrificing the nuance and depth of the outcome?

While we have argued it was a normative advantage of AI Reflectors, the Habermas Machine was tested on groups of 5 people on average. Indeed, training a personalized reward model, as in the Habermas Machine, for each participant is not scalable. Participants must engage in a pre-deliberation phase for data collection, and personalizing a reward model for each participant becomes computationally intractable as their numbers increase. Along the same lines, the Generative Social Choice approach accounted for 100 people—but the generation broke the group down to smaller groups of at most 20 people due to the limited size of the context window (that is, the number of *words* an LM can take as its input) and the LM ability to sense-make of heterogeneous information. There are indeed two primary open questions surrounding scalability: Is it possible to parse thousands of statements simultaneously, or should iterative approaches be developed (and at what cost)? How does the quality of the output change as the length of the input collection of statements increases?

These empirical questions, essential to understanding the trade-offs between the comprehensiveness of the input data and the accuracy of the generated output, remain to the best of our knowledge unanswered.

## *V.2 - Designs to generate syntheses of open-ended reflective thoughts*

The second area of technical research focuses on the methodologies employed to produce collective judgments, analogous to the aggregative step in classical judgment aggregation. Specifically, we wonder: is it possible to augment the traditional use of the will matrix with a focus on discursive analysis that goes deeper than inferred preferences—and what guarantees shall we uphold for collective synthesis to ensure individual equality and fair representation in collective discourse?

While Tessler et al. (2024) employed a reward model (that specializes in number prediction) to estimate the extent to which participants like new statements, more recent approaches use LMs directly to predict human preferences. In fact, Tessler et al. (2024) explored using a prompted LM to produce the will matrix (instead of a personalized reward model), or rather, to directly output a ranking for each person. They suggest this is a promising approach, albeit strictly worse to their original method. Similarly, Fish et al. (2023) used an LM prompt to predict a person's preference for a statement based on their survey responses and a few-shot prompt. By validating their approach with real user ratings, they found a significant association between human rankings and LM-predicted ones. A recent breakthrough in this area, utilizing models with large context windows, demonstrated that "generative agent simulations" could model human behavior with 85% accuracy simply by parsing a multi-hour interview of individuals as prompt (Park et al. 2024). Notably, the authors observed that while prompts filled solely with demographic information, as in Jarrett et al. (2023), led to more racially and ideologically biased outputs, using open-ended surveys filled with rare, distinctive, and profoundly personal anecdotes significantly improved the LM's capacity to respond accurately.

We question, however, the emphasis on the will matrix to make sense of free-form and open-ended opinions. First, technically, it is unclear whether LMs are put to their best use to provide a numbered utility for a given person-statement pair (but we could imagine other kinds of classifiers developed to that end). Second, conceptually, it is unclear whether a will matrix is well suited at all to represent debates (with the complexity of underlying reasons and epistemological frames,[31] of irreconcilable differences and of meaningful consensus). While using a will matrix brings up undeniable advantages (e.g., it can be probed to justify why a certain output statement was created, it was the only method pre-LMs to aggregate free-form texts), we ponder whether there may be an opportunity to provide new (perhaps enhanced) representation of collective judgements with LMs. If indeed, the goal of AI Reflectors is to better reflect a group's plural stories *in context*, shouldn't we also invest in representing the nuanced relationships between experiences, beliefs, opinions and emotions—and how they compose a polity?

Could we use discursive approaches well-explored by social scientists and build intermediary steps on the way to generating output statements?[32] We lack a clear technical proposal to streamline this idea—and discuss at a high-level hereafter what we hope can be seeds for more concrete proposals. Earlier technical work has explored fine-tuning LMs for summarization (Stiennon et al 2020) and training models on relational representation of values (Klingefjord et al 2024).[33] Building on such initiatives, could we derive methods to embed knowledge about e.g., discourse analysis, from surveys thematic analyses (Braun & Clarke 2006), to frame analyses (Benford & Snow 2000), from argument mapping (Rapanta & Walton 2016, Fairclough & Fairclough 2013) to controversy mapping (Latour 2007 Venturini 2012). Would fine-tuning various LMs to excel in each of these tasks help building sound AI reflectors?

The answer to this question may never be fully resolved, as the optimal statement will remain elusive. Nonetheless, studies could statistically analyze how sensitive the LMs' outcomes are to diverse representativeness criteria. For instance, are LM-generated statements more majoritarian or proportional in their summaries? Do LMs possess the understanding necessary to capture the subtleties of dissensus or consensus? To address such questions, statements could be meticulously crafted with unbalanced perspectives, followed by statistical analyses to determine whether LM outputs over- (or under) represents majority views. Additionally, when it comes to usage, a trade-off will likely have to be made between a more informative and nuanced summary and a more engaging, eye-catching one.

[31] See for instance danah boyd's discussion about differences in epistemological frames here https://www.zephoria.org/thoughts/archives/2018/03/09/you-think-you-want-media-literacy-do-you.html
[32] We thank Bernard Reber for this formulation.
[33] See also Jigsaw's sense-making initiative https://medium.com/jigsaw/making-sense-of-large-scale-online-conversations-b153340bda55

## *V.3 - Biases in the output syntheses of open-ended reflective thoughts*

The next component of our research agenda focuses on the propagation of unaccounted for biases through AI summaries, and how these biases may compromise the integrity of AI Reflector outputs. Can we anticipate, mitigate, or at least account for the impact of inherited biases, text homogenization, and political cues on the meaning and usability of AI-generated summaries ?

While AI summaries have been widely appreciated for their efficiency and effectiveness (Stiennon, 2020), we caution against three pressing issues that require urgent attention: the perpetuation of stereotypical biases, the homogenization of text, and the emergence of value systems in LMs. These concerns are particularly relevant in the context of AI Reflectors, where the AI is prompted to summarize human reflective thoughts, potentially amplifying existing biases and distorting the intended meaning.

Research has shown that LMs can reproduce and even amplify biases present in their training data, which are often rooted in societal stereotypes (Kotek et al., 2023). Next, LMs were reported to exhibit a homogenizing effect in what (Sarkar 2024) called a "mechanised convergence" and (Anderson, Shah, and Kreminski 2024) measured as less semantically diverse than human-created alternatives when used for ideation. It is believed that biases in training data cause certain words and phrases to be oversampled, reinforcing dominant linguistic patterns. Recent research has further warned that such trends could get reinforced in the future as models are trained on synthetic data (that is, data produced by AIs), leading to a model's semantic space getting greatly narrowed (Zhang et al. 2024; Dohmatob et al. 2024).

On another note, models have been found to have emergent value systems (Mazeika, Mantas, et al 2025), or, relatedly, a tendency to exhibit political left-wing biases (Potter et al. 2024 elegantly titled their paper "Hidden persuaders"), with a wealth of such results at this point (Feng et al., 2023; Röttger et al., 2024; Motoki et al., 2024 (across languages Hartmann et al., 2023) that worsen during fine-tuning (Santurkar et al. 2023). This is a specific kind of bias we single out due to the application domain of AI Reflectors.

While this research finds biases when models are asked to answer political questions, research is necessary to understand whether these biases would percolate in synthesis generation—or whether the LMs would perform faithfully as summarizers even as they exhibit biases as question answerers.

In both contexts, both qualitative (through expert reviews) and quantitative (through embedding analyses) studies may be needed to investigate the extent to which homogenisation and political cues impact the meaningfulness of the output statements.

### *V.4 - Human-computer interaction in elicitation of reflective thoughts*

Next, we focus on the elicitation of reflective thoughts through human-AI interaction. We ask: How may we design an interaction framework that scaffolds internal deliberation without coercion and manipulation by, or overreliance on, AI?

Recall that we envision the first step of the AI reflectors staging an AI-human discussion where the human may ask factual questions and the AI may share alternative perspectives—dubbed with lived experience. While emerging research has found positive uses for argumentative AIs, much is to be feared in terms of the AI meddling with human thoughts in an "unnatural way" or in terms of the human developing overreliance in its opinion-forming process.

Costello, Pennycook, and Rand (2024) found that conversing with an AI able to "sustain tailored counterarguments and personalize [...] in-depth conversations" could reduce beliefs in conspiracy theories over the long run. In the same vein, (Argyle et al. 2023) found leveraging "AI chat assistant that makes real-time, evidence-based suggestions for messages in divisive online political conversations [...] improve[d] political conversations without manipulating participants' views." Other research finds that AI storytelling may supersede that of humans (Chu and Liu 2024) and investigates avenues to foster critical thinking through AI interactions through what they call, e.g., Socratic AI (Duelen, Jennes, and Van den Broeck 2024), Socratic Minds (Hung et al. 2024), or WisCompanion (Etori and Gini 2024). While we struggle to understand where conversational greatness emerges from in AI, they seem to exhibit emergent pro-social capabilities that may be leveraged to scaffold human thoughts, as a debate club would.

At the same time, legitimate worries arise when thinking about the coercive power of AI over human thoughts in controlling narratives (Jones and Bergen 2024; Tarsney 2025) or enforcing centralized censorship to control narratives through the technology.[34] LMs are known to hallucinate—confidently outputting misleading or outright wrong information (Rawte, Sheth, and Das 2023) and may spread false beliefs during the reflective step. Even if we were to solve the issues of AI's biased answers, we shall still worry about humans' tendency to over-rely on AI systems (Zhai, Wibowo, and Li 2024) and to surrender to a suggestion or an objection provided by an AI.

The sycophantic and subservient behavior of large language models carries a risk of epistemic deference from users that can hardly be overstated[35]. Surrendering democratic deliberation to a dematerialized process, devoid of face-to-face interaction, might ultimately amount to abandoning deliberation itself. Over time, outsourcing collective reasoning in this way risks eroding the experience of vibrant, embodied human disagreement, undermining our capacity for communication and tolerance of dissent. As we attempt to build more reflective AI, whether used in deliberative or educational setups (Tahiru 2021), we should reflect on how to evaluate their impact on humans' capacity to think critically (Larson et al. 2024). If it is not possible to build hallucination-free or bias-free LMs, we are interested in the narrow path whereby the positive impacts of AIs are exploited while humans develop a hyper-aware sense of their flaws.[36]

[34] https://freedomhouse.org/report/freedom-net/2023/repressive-power-artificial-intelligence
[35] We thank Elizabeth F. Cohen for this note of caution.

How may research help identify conditions on the AI infrastructure and the human environment for the narrow path to shape up? Such research evidently sits at the intersection of behavioral science, psychology, and human-computer interactions—and we suggest partnering with human experts (such as facilitators) to help evaluate the coercive impact and overreliance tendencies in lab experiments. A deliberation facilitator's most prominent role is to allow a free-flow, coercion-free conversation—and they are trained to screen for power and argument imbalance. Inspired by qualitative inductive studies as in (Mansbridge et al. 2012), we vouch for a productive intersection of qualitative and quantitative sciences to control for the negative impacts of persuasive AI and help strike a balance between scaffolding critical thinking and convincing.

# VI - Concluding Thoughts

No single institution can be expected to realize the full value of democracy (Mansbridge et al. 2012). We do not see AI reflectors as a new image of what democracy should look like. Rather, we see them as "countervailing" channels of expression, at a time when deep disagreements, elite-driven politics and unequal political influence are sometimes seen as threats to democratic self-government (Bartels 2008, 2023; Bächtiger and Dryzek, 2024).

Would debates on issues such as abortion, immigration, climate change, or the death penalty be as divided if we could access their underlying reasons—free from party identities and inflammatory candidates (Bullock and Lenz 2019)? Although this question is largely speculative, experiences from mini-publics suggest that common grounds and frames—often buried under hyper candidate-focused politics—can re-emerge during reflective discussions (Niemeyer and Dryzek, 2007; Fishkin, Siu, Diamond et al. 2021). In a world not framed as a zero-sum struggle for majority influence, and one that does not presuppose democracy as inherently adversarial, might elements of unitary democracy begin to surface on a larger scale, overcoming the constraints of face-to-face discussions (Mansbridge 1980)? We do not claim that democracy is or should be non-adversarial; rather, we contend that both adversarial and unitary modes exist, each suited to different contexts, although the current landscape predominantly favors adversarial dynamics. We conclude this essay by asking: can AI Reflectors foster unitary moments that build consensus and help make sense of our differences?

If even a fraction of this hypothesis is validated by pointing out meaningful cleavages and consensus statements or articulating subtly balanced collective judgements, it could offer an opportunity for the collective will to better recognize itself. In Habermasian terms, our understanding of the social world would shift across both tracks of deliberative democracy. Open-ended analyzers might not only foster self-awareness and seed collective action in the second track of the political public sphere—that is, as tools for democratic sense-making and self-organisation—but also provide additional information and exert pressure in the first track as bending rather than binding tools. AI Reflectors are dual use: while it may empower citizens to better understand themselves and to increase pressure on governments, motivated leaders might also exploit this information to adapt and post-rationalize their platforms. There is the risk of malicious actors hacking the system to steer outcomes toward predetermined agendas, thus feeding the darker aspects of representation. How can we ensure widespread technological access and independent model providers to prevent mass manipulation?

[36] In fact, public knowledge about hallucinations, as well as first-hand experience thereof, might advantageously counterbalance overreliance on AI's input. From this perspective, paradoxical as it may seem, hallucinations may indeed help safeguard epistemic vigilance and critical thinking.

Ada Palmer's old fridge metaphor evokes an image of an all-solving technology—seemingly free from the flaws of current systems—until, inevitably, it is hacked (Schneier 2023).[37] Although it is impossible to predict the eventual use of AI Reflectors, we recognize that they grant access to a novel kind of social knowledge that we hope to have theorized paying close attention to the underlying technology and the related philosophical formulations. While we acknowledge their potential for positive application, we also caution against the attendant risks. AI Reflectors might serve as a complex "formula for better understanding our wants and caprices." Rather than causing us to "stop wanting," could they instead help us critique and collectively make better sense of our collective aspirations and differences?

[37]https://www.schneier.com/blog/archives/2025/01/third-interdisciplinary-workshop-on-reimagining-democracy-iword-2024.html